\documentclass[conference]{IEEEtran}
\IEEEoverridecommandlockouts
\usepackage{balance}

\usepackage{comment}
\usepackage{booktabs}

\usepackage{tabularx}   % for X columns that auto-stretch
\usepackage{booktabs}   % for \toprule, \midrule, \bottomrule, \addlinespace
\usepackage{array}      % for >{\raggedright\arraybackslash} column modifiers
\usepackage{graphicx}
\usepackage{cite}
\usepackage{amsmath,amssymb,amsfonts}
\usepackage{algorithmicx}
\usepackage{textcomp}
\usepackage{xcolor}

\usepackage{amsthm}

\usepackage[caption=false]{subfig}

\usepackage{authblk}

\newtheorem{theorem}{Theorem}

\usepackage{algorithm,algpseudocode}

\usepackage{multicol}

\usepackage{mathtools}

\usepackage{cuted}

\usepackage{color,soul}
\def\BibTeX{{\rm B\kern-.05em{\sc i\kern-.025em b}\kern-.08em
    T\kern-.1667em\lower.7ex\hbox{E}\kern-.125emX}}

\makeatletter
\newcommand{\linebreakand}{%
\end{@IEEEauthorhalign}
\hfill\mbox{}\par
\mbox{}\hfill\begin{@IEEEauthorhalign}
}

\makeatother

\begin{document}

\title{
Ghost Target Elimination in 5G Reference Signals ISAC via Multi-static Multilateration

\thanks{This work has been supported by the Smart Networks and Services Joint Undertaking (SNS JU) project 6G-DISAC under the EU’s Horizon Europe research and innovation program under Grant Agreement no. 101139130}
}

\author[1]{Ali Al Khansa}
\author[2]{Sami Mekki}
\affil[1]{Orange Labs, Rennes, France}
\affil[2]{Nokia Standards, Massy, France}

\affil[ ]{Emails: ali.alkhansa@orange.com, sami.mekki@nokia.com}

%\affil[ ]{Emails: \{ali.alkhansa;giyyarpuram.madhusudan\}@orange.com, sami.mekki@nokia.com}

\maketitle

\IEEEpubidadjcol

\begin{abstract}
% Verify after finishing the paper and do so as well for the intro
Integrated Sensing and Communication (ISAC) systems can reuse 5G New Radio (NR) reference signals to enable joint communication and environmental sensing. Among these signals, the Positioning Reference Signal (PRS) is particularly suitable for range and Doppler estimation. However, its comb structure introduces periodic ambiguities that generate ghost targets. Existing solutions resolve these ambiguities by combining PRS with additional signals such as the DeModulation Reference Signal (DMRS), which increases sensing overhead. This paper proposes a PRS-only sensing framework that resolves ambiguities using multi-static observations. Instead of removing ghosts locally at each receiver, the proposed method exploits spatial diversity and performs ambiguity resolution at a fusion anchor. The anchor jointly estimates target locations and ambiguity indices via multilateration across multiple sensing nodes. %, jointly estimating target positions and ambiguity indices. 
This approach eliminates the need for additional reference signals while maintaining low receiver complexity. Simulation results show that the proposed method reliably resolves ghost targets and achieves superior detection performance compared to benchmark methods across various scenarios, while significantly enhancing resource efficiency by enabling sensing using PRS-only transmissions.

\end{abstract}

\vspace{-0cm}


\vspace{-15 pt}
\section{Introduction}
% Read and proofread logic and English
% Verify if I missed things I could have added from patent text OR MULTISCOUT TEXT (e.g., multistatic focus)
% Verify the SotA citations
% Verify the contribution part and the rest of the paper
Future 6G wireless networks are expected to extend the role of communication systems beyond data delivery by enabling environmental sensing capabilities within the same infrastructure. This vision is embodied in the Integrated Sensing and Communication (ISAC) paradigm, which supports applications such as object tracking, autonomous driving, environmental monitoring, and smart city services \cite{10944644,strinati2024distributed}. In this context, the reference signals already defined in 5G New Radio (NR) have attracted considerable attention for sensing purposes. Their well-designed structure, favorable correlation properties, and robustness to noise make them appealing for radar-like sensing tasks, while allowing seamless integration with communication functionalities \cite{wei20225g,lyu2025reference}. %Reusing these signals can therefore enable sensing capabilities without requiring additional hardware or major changes to the communication framework.

Several studies have investigated the use of 5G reference signals for sensing \cite{lyu2025reference}. For instance, \cite{wei20225g} compares the sensing performance of Positioning Reference Signals (PRS) with other pilots such as DeModulation Reference Signals (DMRS), Channel State Information Reference Signals (CSI-RS), and Synchronization Signals (SS). Among these, PRS, introduced in 3GPP Release 16, has emerged as a strong candidate due to its flexible configuration, high processing gain, and large time–frequency occupancy, which enable accurate time-of-arrival estimation and extended sensing range \cite{wei20225g}.
Despite these advantages, PRS-based sensing suffers from an inherent limitation caused by its comb-like frequency-domain allocation. The regular spacing between active subcarriers leads to periodic ambiguities in the range domain, resulting in multiple ghost targets \cite{khosroshahi2024leveraging}. These ambiguities significantly degrade detection reliability, particularly in cluttered or multi-target environments, and represent a major obstacle to PRS-only sensing.
%and rich time–frequency resources. %Its pseudo-random sequence structure provides favorable correlation properties for range estimation, often outperforming other reference signals in sensing applications \cite{wei20225g}.
%However, the comb-based allocation of PRS leaves empty subcarriers between active tones, which creates periodic range ambiguities. As a result, multiple ghost targets appear in the range domain, making reliable detection challenging.

Various approaches have been proposed to address this ambiguity problem. A prominent line of work resolves PRS ambiguity by jointly exploiting multiple reference signals \cite{wei2024multiple}. %In particular, the works in 
In particular, 3GPP-compliant solutions combining PRS with DMRS have been proposed to eliminate range or Doppler ambiguities \cite{khosroshahi2024leveraging,khosroshahi2024doppler}. 
%propose 3GPP-compliant solutions that combine PRS with DMRS to eliminate range or Doppler ambiguities. 
Other methods attempt to mitigate ambiguities through irregular resource allocations, such as the irregular time–frequency patterns proposed in \cite{golzadeh2024joint}. Despite their effectiveness, these approaches present several limitations. Leveraging multiple reference signals increases system complexity and sensing overhead, as additional resources must be allocated for sensing tasks. Irregular resource allocation strategies also deviate from standard signal structures and can affect correlation properties, potentially increasing sidelobe levels \cite{hakobyan2019ofdm}. More broadly, allocating additional resources to sensing directly impacts the communication capacity, making resource-efficient sensing strategies challenging \cite{al2024dynamic,al2025adaptive,al2025learning}.

%In this paper, we address the ambiguity introduced by the comb structure while avoiding the need for additional reference signals. Instead of resolving ghost targets locally at each receiver, we exploit spatial diversity in a multi-static sensing architecture. Each sensing node performs PRS-based detection and forwards its bistatic range estimates to a fusion anchor. The anchor then resolves the ambiguity by jointly estimating the target position and the corresponding ambiguity indices through multilateration. This approach enables reliable sensing using PRS-only transmissions while keeping receiver complexity low and preserving communication resources. Numerical results show that the proposed framework accurately resolves ghost targets and achieves superior detection performance compared to benchmark methods, while significantly improving resource efficiency. 

In this paper, we propose a \emph{PRS-only} sensing framework that resolves range ambiguities without requiring additional reference signals or non-standard resource allocations. The key idea is to exploit spatial diversity through a multi-static network of sensing nodes and to perform ambiguity resolution centrally rather than locally. Instead of attempting to eliminate ghost targets independently at each receiver, the proposed approach jointly processes multi-static observations at a fusion anchor, which estimates both the true target locations and the associated ambiguity indices via multilateration. By leveraging geometric consistency across spatially separated sensing nodes, the method effectively suppresses ghost targets while maintaining low receiver complexity. Simulation results demonstrate that the proposed framework reliably resolves PRS-induced ambiguities and significantly outperforms benchmark methods in terms of detection accuracy across a wide range of scenarios. Importantly, this is achieved without increasing sensing overhead, thereby enabling efficient and fully standards-compliant PRS-based ISAC operation.

The rest of the paper is organized as follows: Section II introduces the system model, Section III presents our proposal, Section IV provides the numerical results, and Section V concludes the paper.

%\section{ISAC system Model}
%\rv{Here, I used the same formulations presented in Nokia papers. However, I omitted the description of 5G refernce signals as it is not the main scope of this work. We can add more things later on according to the size limitations of the paper.}
%\hl{KiM the differences between the two papers (notation + last equations) + Braun (the wholse strategy which I have to recheck to understand more). \\
%I used Doppler paper (notation), but then, I used the last equations from the range paper.}
% KiM that there was some differences here between Nokia formulations and Braun's formulation. So, maybe recheck it later on.

% THE ABOVE COMMENTS ARE OLD BUT I KEPT THEM
% NOW, I INTRODUCE VIA MULTISCOUT THEN TALK PER RECEIVER

\section{Multi-Static ISAC System Model}
% REREAD TEXT AND LOGIC
% COMPARE WITH CAMERA READY (TEXT + !!EQUATIONS!!)
% VERIFY IF MORE FROM SCOUT (following what I write in trilat)

\begin{figure}
%\vspace{pt}
\centering
\includegraphics[scale=0.25]{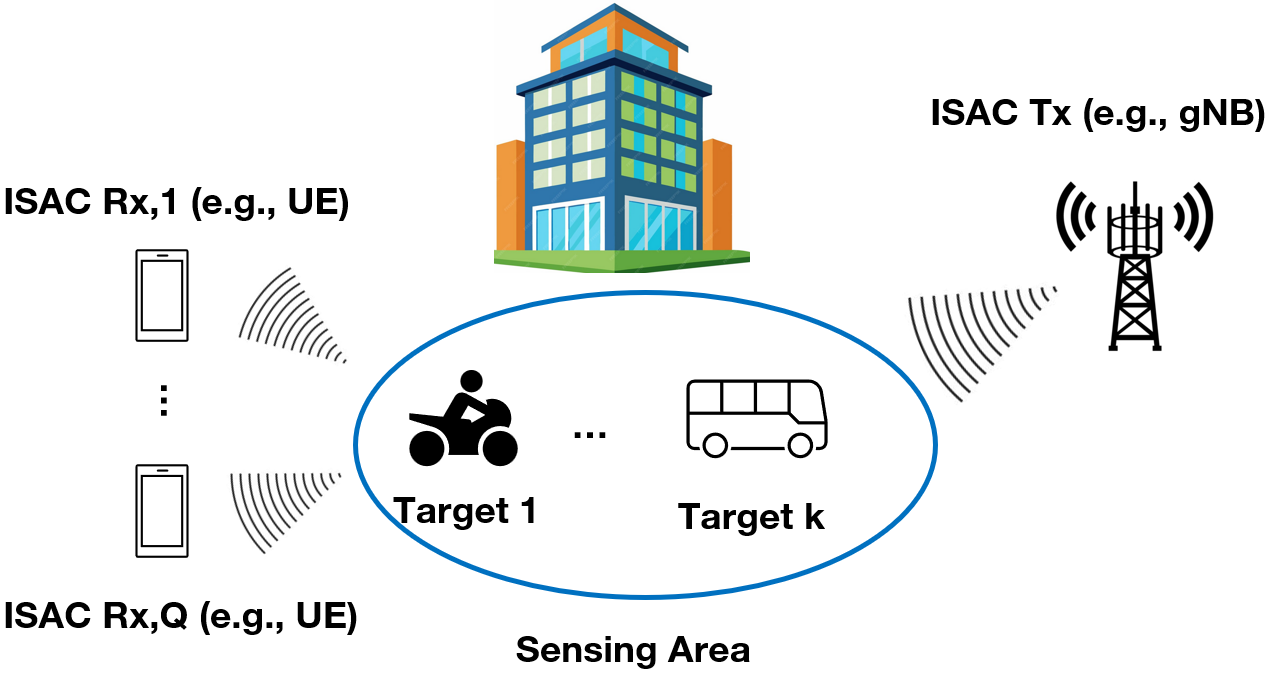}
\caption{Multi-Static ISAC system model.}

\label{fig:SystemModel}
\end{figure}

We consider the multi-static scenario illustrated in Fig.~\ref{fig:SystemModel}, where a 5G gNB acts as the transmitter while $Q$ spatially distributed receivers observe the reflections from the sensing scene with $q \in \{1,\dots,Q\}$ representing the receiver index. The receivers operate in a passive sensing mode and estimate bistatic ranges using the echo received by the transmitted signal. The sensing region contains $K$ point-like targets present in the line-of-sight of both the transmitter and the receivers.

\subsection{OFDM Radar Signal Model}

The gNB transmits an OFDM waveform composed of $N$ symbols in time and $M$ subcarriers in frequency. The transmitted continuous-time signal can be written as \cite{khosroshahi2024leveraging}:
\begin{equation}
s(t) = \sum_{n=0}^{N-1} \mathrm{rect}\Bigl(\tfrac{t-nT_0}{T_0}\Bigr)
\sum_{m=0}^{M-1} d(m,n) e^{j2\pi m\Delta f (t-nT_0)},
\end{equation}
where $\Delta f$ denotes the SubCarrier Spacing (SCS), $\mathrm{rect}(\cdot)$ denotes the rectangular pulse, $T_0 = T_{\text{CP}} + T_s$ is the OFDM symbol duration with $T_s=1/\Delta f$ and $T_{\text{CP}}$ the cyclic prefix length, and $d(m,n)$ represents the symbol transmitted on subcarrier $m$ of OFDM symbol $n$.
%Assuming $K$ point targets in the sensing region, 
The signal received at receiver $q$ can be modeled as \cite{braun2014ofdm}:
\begin{equation}
r_q(t) = \sum_{k=1}^{K} \beta_{k,q} s\bigl(t-\tau_{k,q}\bigr)e^{j2\pi f_{d,k,q} t} + u_q(t),
\end{equation}
where $\beta_{k,q}$ denotes the reflection coefficient, $\tau_{k,q}$ is the bistatic propagation delay associated with the bistatic path, $f_{d,k,q}$ is the Doppler shift, and $u_q(t)$ represents additive white Gaussian noise with zero mean and variance $2\sigma^2$. After sampling and applying the Fast Fourier Transform (FFT) at the receiver, the received sample on subcarrier $m$ and symbol $n$ is written as:
\begin{equation}
\begin{aligned}
y_q(m,n) &= \sum_{k=1}^{K} \beta_{k,q}
e^{j2\pi nT_0 f_{d,k,q}}
e^{-j2\pi m\Delta f \tau_{k,q}} v(m,n) \\
&\quad + \mathrm{ISI}_q + z_q(m,n),
\end{aligned}
\end{equation}
where $z_q(m,n)$ denotes the complex AWGN component with zero mean and variance $2\sigma^2$. As in conventional OFDM radar processing, uncompensated delays or Doppler shifts may introduce Inter-Symbol Interference (ISI) under challenging sensing conditions.

\subsection{Bistatic Range Estimation}

Each receiver estimates the bistatic ranges of potential targets using a periodogram-based approach due to its low computational complexity. The received samples $y_q$ are first divided element-wise by the transmitted symbols $d$ to obtain a normalized matrix $g_q$. Range estimation is performed by applying an $M$-point Inverse FFT (IFFT) along the frequency dimension, yielding $s_{n,q}(l)$ for each OFDM symbol. Averaging across the $N$ symbols improves the signal-to-noise ratio, producing $\overline{s}_q(l) \triangleq \frac{1}{N}\sum_{n=0}^{N-1} s_n(l)$.
The peak index $\widehat{l}_{k,q}$ corresponds to the estimated bistatic range $\widehat{B}_{k,q} = \frac{c\,\widehat{l}_{k,q}}{\Delta f\,M}$. Similarly, Doppler estimation can be obtained after delay compensation by applying an FFT along the time dimension \cite{braun2014ofdm}. However, in this work, the primary measurements used for localization are the bistatic ranges estimated at each receiver. These measurements are then forwarded to the fusion anchor, where spatial processing is used to resolve ambiguities and estimate the target positions. For brevity, detailed derivations are provided in \cite{khosroshahi2024leveraging}.

\section{A PRS-Only Ambiguity Elimination Proposal}
% Subsections A, B, and C should be paraphrased if I want to reduce exact text from ICC paper

%In this section, we start by describing the ghost target problem and one benchmark solution of using PRS with DMRS to remove them. Then, we describe our proposed ghost elimination via the multilateration method for newly arrived targets in the considered multi-static scenario. After that, we present our complete proposal that uses successive estimation with a matching process for cases of no newly arrived targets.
This section addresses the ghost target problem inherent to PRS-based sensing and introduces a PRS-only ambiguity elimination framework suitable for multi-static ISAC deployments. We first briefly review a benchmark solution that combines PRS and DMRS to suppress ambiguities. We then present the proposed multilateration-based approach, which exploits spatial diversity to eliminate ghost targets without relying on additional reference signals.

\subsection{Ghost Targets and How to Use DMRS to Remove Them}
\begin{figure}[t!]
%\vspace{-5 pt}
\centering
\includegraphics[scale=0.3]{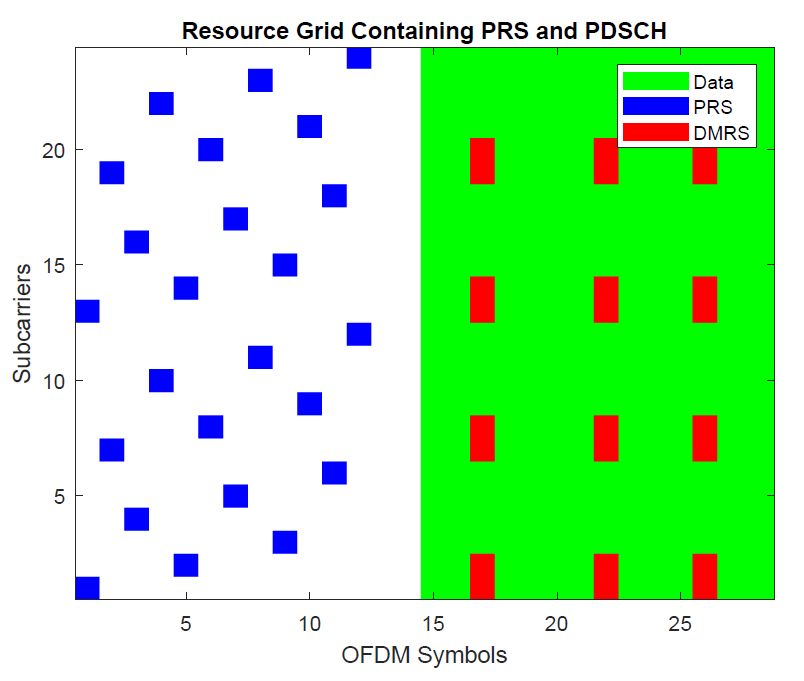}
\caption{PRS–PDSCH resource allocation: 2 slots, 2 PRBs, comb size 12 \cite{khosroshahi2024leveraging}.}
\label{PRS&DMRS}
\end{figure}

When used for passive target sensing, the comb-structured allocation of the PRS in 5G NR introduces empty subcarriers that cause periodic ambiguities (replicas of the real targets) in the range and Doppler domains. The work of \cite{khosroshahi2024leveraging} proposes removing these ghosts by jointly exploiting the PRS and %via using 
the DMRS embedded in the Physical Downlink Shared Channel (PDSCH), which is  repurposed for sensing, as presented in Fig. \ref{PRS&DMRS}. %%%illustrates an example of such resource allocation in the time domain between PRS and PDSCH, and shows an example of the comb-structure of PRS where  $K_{\text{comb}}=12$. %when comb size is 12.
Since PRS and DMRS exhibit different frequency-domain structures, their corresponding ambiguity patterns differ. By appropriately combining their sensing outputs, ghost targets can be suppressed while preserving the true target response.
%The idea is to exploit the fact that PRS and DMRS have different ghost patterns and by combining their outputs appropriately, ghost peaks might get suppressed. %, leaving only the real targets. %%%In this approach, PRS is first processed using standard range estimation to produce a range profile, and the same is done for DMRS extracted from PDSCH. 
Different fusion strategies can be applied such as for range ambiguities \cite{khosroshahi2024leveraging}: % and below are some examples of how to remove range ambiguities \cite{khosroshahi2024leveraging} for $K_{\text{comb}} \in\{2,4,6,12\}$.
\begin{itemize}
    \item For smaller comb sizes (2, 4), %the ghost patterns of PRS and DMRS are sufficiently distinct. Here, 
    the range profiles from the two signals are combined using a Hadamard product since the ghost in one profile typically corresponds to a low value in the other.%, this product cancels ghosts while preserving real peaks.
    \item For larger comb sizes (6, 12), %some ghost peaks overlap between PRS and DMRS, so direct multiplication is less effective. In this case, 
    the method sums the time-domain representations of PRS and DMRS, thereby disturbing the periodicity of the ghost pattern and reducing the magnitude of ghost peaks below detection thresholds.
\end{itemize}
This solution maintains range resolution and maximum detection range and the two cases (PRS-only and PRS+DMRS) are illustrated in Fig. \ref{fig:OldResults} and will be discussed in section \ref{sectionNR}. %For Doppler ambiguities, a similar method can be applied \cite{khosroshahi2024doppler}. These methods will serve as the initialization step of our proposed algorithm, as detailed in the following subsections.
Despite its effectiveness, the joint PRS–DMRS approach increases sensing overhead and receiver complexity, as it requires allocating additional reference signals and performing multiple sensing chains. These limitations motivate the PRS-only solution proposed next.
%While effective, such approaches face the drawbacks of increased complexity, performance degradation, and higher sensing overhead from allocating multiple reference signals. These limitations motivate our proposal presented next. % that minimizes the use of multiple reference signals while still eliminating ambiguities. %In the following, we introduce a successive estimation strategy designed to achieve this balance, freeing more resources for communication and reducing sensing overhead.

\subsection{Ghost Elimination via Multilateration}
\label{subsectionGhostEliminationviaMultilateration}

%\rv{The aim of this paper is to resolves ghost targets with no reliance on multiple reference signals. ADD TEXT FROM PATENT HERE + TAKE SOME OF THE FOLLOWING SUBSECTION (rearrange)}.

We propose a PRS-only ambiguity elimination strategy that leverages spatial diversity in a multi-static sensing configuration. Instead of attempting to remove ghost targets locally at each receiver, the proposed method resolves ambiguities centrally at a fusion anchor by jointly processing multi-static bistatic range measurements.

%We propose a strategy that removes ghost targets for each newly appearing target. 
%Although we have ghost targets in the bistatic distances at each of the receivers (e.g., Fig \ref{fig:OldResults}), we can make use of the multiple receivers and remove ghosts at the fusion anchor. This is possible due to the periodic patterns that ghost targets appear in, and which can be exploited to eliminate them. For a given $K_\text{comb}$, with $c$ representing the speed of light, the ghost targets of each bistatic distance appear at multiples of $D$ 
Due to the comb-based structure of PRS, ghost targets appear periodically in the bistatic range domain (e.g., Fig \ref{fig:OldResults} top plot). For a given comb size $K_{\text{comb}}$, the spacing between successive ghost replicas is given by 
\cite{golzadeh2024joint}:
\begin{equation}
    D = \frac{c}{K_{\text{comb}} \Delta f},
\label{Dghost}
\end{equation}
where $c$ denotes the speed of light and $\Delta f$ is the subcarrier spacing. As a result, each receiver observes multiple candidate bistatic ranges corresponding to the true target and its ghost replicas.\\
In practical ISAC systems, the ultimate objective is not the bistatic range itself but the estimation of the target position. This motivates a shift from local ambiguity resolution to position-domain processing. In the proposed framework, bistatic range estimates from multiple spatially distributed receivers are forwarded to a fusion anchor, which performs multilateration to estimate the target location \cite{widdison2024review}.
While ghost targets act as ambiguities at individual receivers, they do not exhibit geometric consistency across multiple sensing nodes. By extending the multilateration formulation to account for ambiguity indices associated with each bistatic measurement, the fusion anchor can jointly estimate the true target position and discard geometrically inconsistent ghost hypotheses. This enables reliable ghost suppression without modifying the PRS waveform or allocating additional reference signals.

The proposed approach is particularly well suited for scenarios in which new targets appear sequentially, allowing ambiguity resolution to be performed at each update. In the following subsection, we extend this principle to more general scenarios using a successive estimation and matching strategy.
\subsection{Multi-static Multilateration with Ambiguity}
We consider a multi-static sensing configuration composed of one transmitter and $Q$ spatially distributed receivers observing a common target. Each receiver estimates a bistatic range based on PRS processing. Due to the comb-based structure of PRS, these bistatic range estimates are affected by periodic ambiguities, such that multiple candidate ranges are observed for a single physical target.

Let $B_{q}$ denote the bistatic range between the transmitter, the target, and receiver $q$. At a given receiver $q$, and due to the presence of ghost targets, $B_{q}$ can be expressed as
%Due to the presence of ghosts, a receiver $q$ will provide the fusion anchor with an ambiguous estimate of the bistatic range of the newly arrived target at position $\mathbf{p}$. Analytically, a cluster of possible estimates can be written as:
\begin{equation}
B_{q} = b_q + n_q \cdot D,
\label{eq:ambiguous_range}
\end{equation}
where $n_q \in \mathbb{N}$ is the ambiguity index associated with receiver $q$, $D$ is the ambiguity period defined in \eqref{Dghost}  and $b_q$ is the minimum peak value (whether real target or ghost target). 
Each receiver therefore provides a set of candidate bistatic ranges corresponding to different values of $n_q$, which can be interpreted as a family of ellipses in the spatial domain. For receiver $q$, the foci of these ellipses are given by the transmitter and the receiver locations, while the bistatic range $B_q$ determines the length of the corresponding major axis.
%Thus each measurement therefore, defines a family of ellipses corresponding to different ambiguity hypotheses.
Rather than resolving these ambiguities independently at each receiver, the proposed approach exploits spatial diversity by performing joint processing at a fusion anchor. The key observation is that while ghost targets are locally indistinguishable from true targets in the range domain, they do not satisfy the geometric consistency required across multiple sensing nodes.
%The goal of multilateration is 
This motivates an ambiguity-aware multilateration formulation, whose objective is to  jointly estimate the target position defined by $\mathbf{p} \in \mathbb{R}^2$ (or $\mathbb{R}^3$), and the ambiguity indices $n_q$ using measurements from multiple receivers. For a given tuple $\mathbf{n } =(n_1,\dots,n_q)$, our proposed multilateration with ghosts can be formulated as:
\begin{equation}
\label{multilaterationequation}
\min_{\mathbf{p},\mathbf{n}} \sum_{q=1}^{Q} 
\left( \| \mathbf{p} - \mathbf{p}_{\mathrm{Tx}} \| + \| \mathbf{p} - \mathbf{p}_{\mathrm{Rx}_q} \| - B_q \right)^2.
\end{equation}
where $\mathbf{p}_{\text{Tx}}$ and $\mathbf{p}_{\mathrm{Rx}_q}$ denote the known positions of the transmitter (Tx) and receiver (Rx) $q$, respectively.

By leveraging multiple independent measurements, the correct combination of ambiguity indices leads to a consistent geometric intersection. %In case of no ghosts, then $n_q = 0$ and the problem reduces to classical multi-static multilateration (check the right plot of Fig. \ref{fig:NR_BlockDiagram}). 
In the absence of ambiguities, i.e., when $n_q = 0$ for all $q$, the problem reduces to conventional multi-static multilateration, as seen in the right plot of Fig.~\ref{fig:NR_BlockDiagram}.
%In contrast, when ambiguities are present, this results in a family of ellipses for each receiver (instead of 1), and the additional integer variables $n_q$ must be estimated jointly with the target position to determine the correct ellipse. %(as illustrated in Fig. \ref{fig:OldResults}). %The redundancy provided by multiple receivers enables resolving these ambiguities and recovering the true target location.

In contrast, when ambiguities are present, each receiver gives rise to a family of candidate ellipses rather than a single one. In this case, the integer-valued ambiguity indices $n_q$ must be jointly estimated together with the target position in order to identify the geometrically consistent ellipse corresponding to the true target.

Except in edge cases, the proposed method reliably distinguishes true targets from ghost targets.
%our method correctly identifies true targets from ghosts. 
However, it increases the computational complexity of the multilateration step, which now scales with the total number of integer combinations across all possible ellipses, given by $\prod_{q=1}^{Q} N_q$, where $N_q$ is the maximum index of $n_q$ for receiver $ q $.
Fortunately, in practice, $N_q$ is naturally bounded by the size of the region of interest, making the approach computationally feasible. Nevertheless, a key limitation arises in multi-target scenarios, where correctly associating measurements across different receivers becomes challenging.
To address both the association problem and the overall computational cost, we propose a hybrid strategy: multilateration with ghost elimination is applied only to newly detected targets, while the successive estimation method is used for already tracked targets, as detailed in the following section.
%However, it increases the complexity of the multilateration process following the total number of integer combinations of the possible ellipses $\prod\limits_{q=1}^{Q}N_q$, with $N_q$ representing the maximum index of $n_q$. Fortunately, in practice, $N_q$ is bounded following the region of interest, thus making the approach computationally feasible. Besides that, one limitation of the proposal is in the case of multiple targets, and the issue of the association between these targets from the different receivers. To solve this, and also to reduce the complexity cost of this method, we propose to use the multilateration with ghosts only for new targets, while using the successive estimation method for targets previously detected, as explained below.

\subsection{Successive Estimation}
\label{subsec:successive_estimation}
While ambiguity-aware multilateration enables reliable ghost elimination, its computational complexity increases with the number of ambiguity hypotheses and becomes more challenging in multi-target scenarios due to data association across receivers. To address these limitations, we propose a successive estimation strategy that exploits temporal consistency between consecutive sensing updates.
The key idea is to apply the ambiguity-aware multilateration procedure only when a new target is detected. Once the position of a target has been estimated and validated, its subsequent updates are tracked using a successive estimation approach that relies on the previously estimated position as a prior. This prior information significantly constrains the set of admissible ambiguity indices, thereby reducing both the search space and the risk of incorrect data association.

The successive estimation process consists of two main components:
\begin{itemize}
    \item \textbf{Peak Clustering:} Upon receiving a new echo, each receiver detects all peaks in the bistatic range profile. These peaks are then clustered based on the predicted ghost spacing $D$, such that all peaks separated by integer multiples of $D$ are grouped into the same cluster. Each cluster therefore corresponds to a set of candidate measurements associated with a single physical target (e.g., two such clusters are illustrated in Fig.~4).

    \item \textbf{Matching Process:} For each target stored in the historical database, the closest detected peak within the corresponding cluster is selected as the true measurement, while the remaining peaks in that cluster are classified as ghost targets. This matching step also enables the identification of targets that have exited the sensing region, as well as newly appearing targets, which are subsequently processed using the ambiguity-aware multilateration procedure.
\end{itemize}

The peak clustering procedure is simple and computationally efficient. First, all peaks are detected from the bistatic range profile and grouped according to their separation by integer multiples of the ambiguity period $D$. Specifically, clustering starts from the minimum detected peak value, and all other peaks whose separation from it is an integer multiple of $D$ are assigned to the same cluster. This procedure is repeated iteratively for the remaining unassigned peaks until all detected peaks are clustered.
Analytically, consider a set of detected peaks with bistatic range estimates $\{B_i\}_{i=1}^{N_{\text{peaks}}}$. The first cluster contains any peak $j \in \{1,\dots,N_{\text{peaks}}\}$ such that the offset ratio  given by
%for a set of peaks with bistatic estimates $B_i$, with $i\in{\{1,\dots,N_{\text{peaks}}\}}$, the first cluster will contain any peak $j\in{\{1,\dots,N_{\text{peaks}}\}}$ such that the offset ratio $\text{OffsetRatio}_j$ is sufficiently close to an integer (e.g., within a predefined threshold $\epsilon$), where:
\begin{equation}
\text{OffsetRatio}_j =   \frac{ B_{j} - \min(B_{i}) }{ D },  \quad i\in{\{1,\dots,N_{\text{peaks}}\}}
\label{OffsetRatio}
\end{equation}
is sufficiently close to an integer, within a predefined threshold $\epsilon$. After forming the first cluster, its elements are removed from the set $\{B_i\}$, and the same procedure is applied recursively to construct the remaining clusters.

%The remaining clusters will be formulated exactly the same after removing from $B_i$ the previously clustered peaks. %The top plot of Fig. \ref{fig:OldResults} illustrates the clustering of peaks process.
Once clustering is completed, the matching process between the detected clusters and previously stored (historical) target estimates is performed. For each historical bistatic range estimate $B_{\text{history},k}$, with $k \in \{1,\dots,N_{\text{history}}\}$, the cluster containing the closest detected peak is selected. The peak corresponding to this minimum distance is identified as the true target measurement, while the remaining peaks within the same cluster are classified as ghost targets and discarded. This matching process allows ghost elimination for previously detected targets without invoking ambiguity-aware multilateration.

Formally, a historical estimate $B_{\text{history},k}$ is matched to the detected peak $B_{\widehat{i}}$, where
\begin{equation}
\widehat{i} = \arg \min_{i \in \{1,\dots,N_{\text{peaks}}\}} 
\left| B_{\text{history},k} - B_i \right|,
~~ \forall k \in \{1,\dots,N_{\text{history}}\}.
\label{matchedpeak}
\end{equation}

%After the clustering, the matching process between the cluster of targets and the previously known (historical) targets starts by selecting, for each historical range estimate, the closest cluster from the current detections. 
%The closest cluster is the cluster that include the closest peak to the historical value. This  peak will be selected as the real target of the cluster, while the other peaks of this cluster are eliminated as they are considered ghosts. Using this, we could eliminate ghosts of previous estimates without the need of performing multilateration with ghosts, as ghosts are already eliminated via the matching process. Analytically, a given historical value $B_{\text{history},k}$, $k\in\{1,\dots,N_{\text{history}}\}$, will be matched to the cluster that includes the peak $B_{\widehat{i}}$, where:
%\begin{equation}
%    \quad \widehat{i} = \arg \min_{i \in \{1, \dots, N_{\text{peaks}}\}} | r_{\text{history},k} - B_i |, \forall k \in \{1, \dots, N_{\text{history}}\}.
%    \label{matchedpeak}
%\end{equation}

\begin{figure}[t!]
\vspace{-0 pt}
\centering
\includegraphics[width=0.48\textwidth]{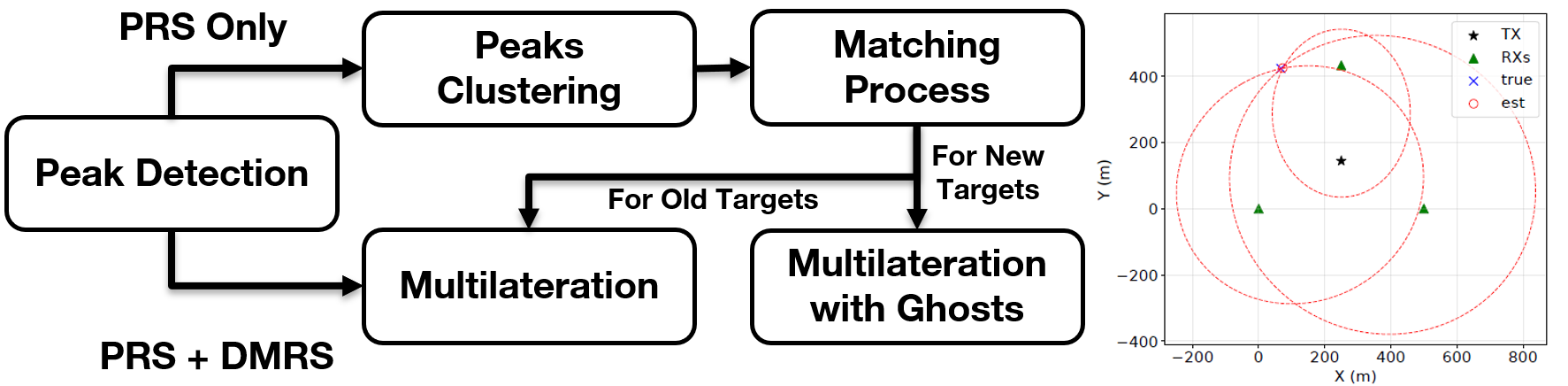}
\caption{The Block diagram of the proposal and a trilateration illustration.}
\label{fig:NR_BlockDiagram}
\end{figure} 

If no detected cluster lies within a predefined tolerance of a given historical estimate—where the tolerance may be chosen based on the maximum expected target velocity—the corresponding target is removed from the history, indicating that it has exited the sensing region. Conversely, any detected cluster that is not associated with a historical estimate is declared as a new target and is processed using the ambiguity-aware multilateration method described in Subsection~\ref{subsectionGhostEliminationviaMultilateration}.

%If a historical value is not close enough to any of the clusters (to given a tolerance value (e.g., tolerance based on the velocity of the targets)), we remove this target from the history as this means this target is no longer in the sensing area. On the other hand, for a cluster of targets that was not matched with any of the historical values, this target is declared as a new target, and thus, we initiate the process of multilateration with ghosts as explained in the previous subsection. 

%The right side of Fig. \ref{fig:NR_BlockDiagram} illustrates one example of multilateration with no ghosts. With ghosts, the method needs to repeat its strategy taking into account different possible ellipses following the indices $n_q$. 
%The left side of Fig. \ref{fig:NR_BlockDiagram} illustrates our proposal and the SotA solution. After detecting the peaks, multilateration is directly called for the SotA benchmark. For our proposal, our algorithm starts by clustering the peaks, and then, the matching process. After that, we update the history, and we perform a normal multilateration with no ghosts (as the SotA benchmark) for old targets and multilateration with ghosts as explained in subsection \ref{subsectionGhostEliminationviaMultilateration} for new targets. That said, the complete process is presented in Algo. \ref{alg:multistatic_algo}. 
The left-hand side of Fig.~\ref{fig:NR_BlockDiagram} illustrates the proposed processing flow alongside the State-of-the-Art (SotA) benchmark. In the benchmark approach, multilateration is directly applied to all detected peaks. In contrast, the proposed method first performs peak clustering and matching, followed by history updating. Standard multilateration without ambiguities is then applied to previously detected targets, while ambiguity-aware multilateration is invoked only for newly detected targets. The complete processing chain is summarized in Algorithm~\ref{alg:multistatic_algo}.

%\vspace{-10 pt}
\begin{algorithm}[ht]
\caption{The Multi-static PRS-Only Sensing Algorithm}
\label{alg:multistatic_algo}
\begin{algorithmic}[1]

    \State $Q$ receivers perform sensing using PRS only and share detected peaks with the fusion anchor.
    
    \State \textbf{Peaks Clustering:} The fusion anchor clusters the peaks following Eq. (\ref{OffsetRatio}).
    \State \textbf{Matching Process:} The fusion anchor matches the historical values with the targets' clusters following Eq. (\ref{matchedpeak}).
    \If{(\textit{There is a new target})} 
        \State Perform multilateration with ghosts to jointly estimate $\mathbf{p}$ and $\mathbf{n}$ for the new target following Eq. (\ref{multilaterationequation}).
        \State Using $\mathbf{n}$, update the history values for the $Q$ receivers.
    \EndIf
    \If{(\textit{There is one or more old targets})} 
        \State Perform standard multilateration (i.e., Eq. (\ref{multilaterationequation}) for $n_q=0$) to estimate $\mathbf{p}$ using the validated bistatic distances.
    \EndIf

\end{algorithmic}
\end{algorithm}

%\vspace{-10 pt}
\subsection{Sufficient Conditions of the Matching Process}

We now examine the practical conditions under which the proposed matching strategy reliably identifies real targets. For a given target previously estimated, the matching process may result in an error if a target moves significantly between two sensing intervals. %In particular, a mismatch may occur when a target shifts far enough toward the location of one of its ghost replicas. 
In such cases, the matching algorithm may associate the target with a previous position rather than its new shifted location. The likelihood of such errors depends the distance between ghost peaks $D$, the velocity of the target $v$, the direction of movement, and the sensing refresh rate, denoted $\Delta t$. Intuitively, the longer the system waits before refreshing the sensing data, or the faster the target moves, the higher the risk of mismatch. To address this, we define a worst-case scenario and derive a sufficient condition on the maximum allowable target velocity that ensures reliable matching. 
This condition enables us to select optimal values for $K_{\text{comb}}$ and the sensing refresh rate $\Delta t$, which are two critical degrees of freedom in balancing the efficiency and robustness of the ISAC system.
\begin{theorem}\label{th1}
Let $v$ denote the target’s velocity. The matching process correctly identifies the real targets if:
\begin{equation}
\Delta t K_{\text{comb}} < \frac{c}{4 v \Delta f}.
\label{SufficientCondition}
\end{equation}
\end{theorem}

\begin{proof}
Over one sensing interval $\Delta t$, the worst-case two-way range displacement is
\begin{equation}
\Delta d \;\le\; 2\,v\,\Delta t.
\end{equation}
Reliable matching requires the displacement to be smaller than half of the distance between adjacent ghost peaks (two-way):
\begin{equation}
2\,v\,\Delta t \;<\; D/2.
\end{equation}
Substituting $D$ and rearranging yields \eqref{SufficientCondition}.
\end{proof}
Theorem \ref{th1} expression offers a practical design guideline. For a given maximum target velocity $v$ and subcarrier spacing $\Delta f$, the system designer can jointly choose the largest possible values of $K_{\text{comb}}$ and $\Delta t$ that satisfy the inequality. Larger values are preferred as they reduce the frequency of sensing operations and the reference signal overhead, thus achieving better resource utilization. Table \ref{TableVel} illustrates the maximum distance (i.e., $v\Delta t$) a target should not exceed with respect to the different $K_{\text{comb}}$ and SCS. An analysis to the effect of the target velocity will be discussed in section \ref{sectionNR}.% in Fig. \ref{fig:PdvsV}).

\begin{table}[H]
    \centering
    \caption{Maximum allowable target displacement $v\Delta t$ derived from the sufficient condition in Eq.~(\ref{SufficientCondition}) for different PRS comb sizes $K_{\text{comb}}$ and subcarrier spacings (SCS).%Displacement threshold (Eq. (\ref{SufficientCondition})) for different $K_{\textbf{comb}}$ and SCS.
    }
    \begin{tabular}{lcccc}
        \toprule
        & \textbf{comb 2} & \textbf{comb 4} & \textbf{comb 6} & \textbf{comb 12} \\
        \midrule
        \textbf{SCS = 60 kHz}  & 614 & 312 & 208 & 104 \\
        \textbf{SCS = 120 kHz} & 312 & 156 & 104 & 52 \\
        \bottomrule
    \end{tabular}    
    \label{TableVel}
\end{table} 

%\subsection{Advantages of the Proposal}
A key advantage of this threshold (Eq. (\ref{SufficientCondition})) is its ability to significantly reduce sensing overhead. % potential to make it possible to significantly reduce sensing overhead. 
By enabling the use of higher values for both $K_{\text{comb}}$ and $\Delta t$, while taking into account the velocity of the target, the system can allocate more resources for %to
communication and reduce the frequency of sensing operations. Assuming that a given baseline is using a fixed values of $K_{\text{comb}}$ and $\Delta t$, and that our method selects a more efficient configuration using $\max(K_{\text{comb}})$ and $\max(\Delta t)$ (if the target's velocity allows), the relative gain in resource efficiency can be expressed as:
\begin{equation}
\text{Gain} = \left[1 - \frac{ K_{\text{comb}} \cdot \Delta t}{\max(K_{\text{comb}}) \cdot \max(\Delta t)} \right] \times 100.
\label{GainEq}
\end{equation}
This simplified expression compares the average number of sensing resource slots used per unit time in our method versus the baseline, assuming that the only considered metrics here are $v$ and $\Delta t$. A higher gain indicates better resource utilization and more capacity left for communication. Table \ref{tablegain} illustrates the achievable gain for $\Delta t \in \{20, 80, 160\}$ ms and different $K_{\text{comb}}$. %Note that $\texttt{Pattern\_Deviation()}$ is a simple Boolean function that returns 1 (true) if there is a pattern deviation detection.

\begin{table}[h]
    \label{alg:range_estimation}
    \centering
    \caption{Percentage gain (Eq. (\ref{GainEq})) in resource usage achieved by the proposed method compared to SotA baseline, for different $\Delta t$ and $K_{\textbf{comb}}$.}
    \begin{tabular}{lcccc}
        \toprule
        & \textbf{comb 2} & \textbf{comb 4} & \textbf{comb 6} & \textbf{comb 12} \\
        \midrule
        \textbf{$\Delta t = 20$}  & 97.9\% & 95.8\% & 93.75\% & 87.5\% \\
%        \textbf{$\Delta t = 40$}  & 95.8\% & 91.6\% & 87.5\% & 75\% \\
        \textbf{$\Delta t = 80$}  & 91.6\% & 83.3\% & 75\% & 50 \\
%        \textbf{$\Delta t = 120$}  & 87.5\% & 75\% & 62.5\% & 25 \\
        \textbf{$\Delta t = 160$}  & 83.3\% & 66.6\% & 50\% & 0 \\
        \bottomrule
    \end{tabular}    
    \label{tablegain}
\end{table}

\section{Numerical Results}
\label{sectionNR}
In this section, we evaluate the performance of the proposed PRS-only multi-static sensing framework and compare it against the SotA benchmark based on joint PRS–DMRS processing. The numerical results aim to assess the effectiveness of ghost target elimination, and the robustness of the proposed successive estimation strategy under target motion.

%The numerical results aim to assess three key aspects: i) the effectiveness of ghost target elimination, ii) the robustness of the proposed successive estimation strategy under target motion, and iii) the resulting sensing resource efficiency gains.

%In thi section, we evaluate the performance of the proposed method 
All results are obtained using MATLAB simulations and its 5G toolboxes, ensuring compatibility with the 3GPP standards. A single gNB transmitting a 5G NR signal with a carrier frequency of 28 GHz and an SCS of 120 kHz toward a sensing area containing multiple targets, and three UEs are used to perform bistatic range estimation. Multiple targets are considered, with targets being introduced sequentially until restarting when reaching a predefined maximum number (check table \ref{tab:positions}). The receiver processes the measurements either using the benchmark fused method, i.e,  PRS+DMRS,  or the proposed PRS-only method. %method with clustering, multilateration for newly detected targets, and successive estimation based on historical information for tracked targets. 
\begin{table}[h]
\centering
\caption{Positions of the transmitter, receivers, and candidate targets}
\begin{tabular}{c c}
\hline
\textbf{Node} & \textbf{Position (x, y, z) [m]} \\
\hline
Tx (gNB) & (0, 0, 0) \\
%\hline
Rx$_1$; Rx$_2$; Rx$_3$ & (250, 200, 0); (450, 150, 0); (178, 405, 0) \\
%\hline
Target$_1$ to Target$_3$  & (-310, 360, 0); (600, 500, 0); (-100, 433, 0) \\ %; (663, 488, 0); (-250, 412, 0) \\

\hline
\end{tabular}
\label{tab:positions}
\end{table} 

The PRS is configured to occupy 12 symbols. The DMRS uses configuration type 2, with three DMRS symbols located at symbol positions {3, 8, 12} within each slot. Additionally, the PDSCH employs mapping type "A". The code rate used to determine the transport block sizes is set to 490/1024, as specified in \cite{3gpp2020nr}. The Virtual Resource Block (VRB) bundle size is 4, and VRB to PRB interleaving is disabled, and we consider the 16 Quadrature Amplitude Modulation (16-QAM) for the PDSCH. 

%To ensure a fair evaluation of the matching procedure, the simulation is reset whenever an error occurs, thereby avoiding error propagation across time.

%The setup involves a gNB transmitting a 5G NR signal with a carrier frequency of 28 GHz and an SCS of 120 kHz towards a sensing area containing two passive targets. The targets are located at bistatic distances of 1057 m and 1242 m, with SNR = 0 dB. 

%Fig. \ref{fig:OldResults} provides an illustrative comparison between the received echo when using PRS only and when using PRS+DMRS benchmark method. The bottom figure corresponds to the PRS+DMRS approach, where the combination of both signals effectively suppresses most of the ghost components and yields dominant peaks at the true target positions. Nevertheless, small residual peaks can still be observed at distances spaced by $D$, which may lead to incorrect detections when using threshold-based peak selection. The top figure shows the PRS-only case, where the periodic structure of the ghosts is clearly visible. Instead of removing these components, our proposal groups them into clusters corresponding to each target and exploits their structure. Our method will then identify new targets and old targets, and then perform multilateration with ghosts and without ghost respectively.

\begin{figure}[t]%{0.329\textwidth}
        \centering
        \includegraphics[width=\linewidth]{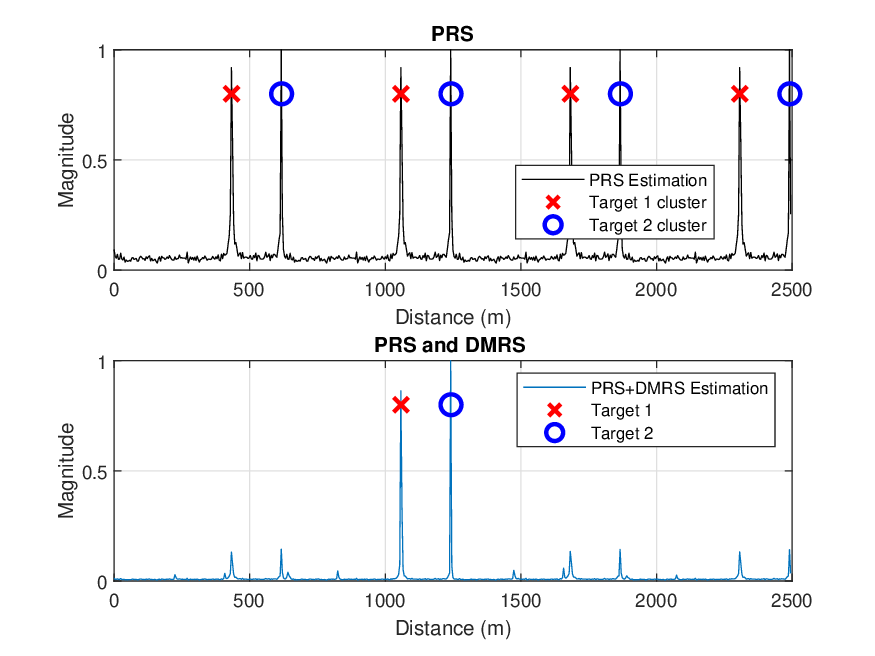}
        \caption{Range estimation with PRS only (top) and PRS + DMRS (bottom) for a two-target scenario ($K_{\text{comb}}$= 4, $f_c$ = 28GHz and SCS = 120kHz).}
        \label{fig:OldResults}
\end{figure}

Figure \ref{fig:OldResults} provides an illustrative comparison between the received echo when using PRS only and the benchmark PRS+DMRS method. The bottom plot corresponds to the PRS+DMRS approach, where the combination of both signals effectively suppresses most of the ghost components and yields dominant peaks at the true target positions. Nevertheless, small residual peaks can still be observed at distances spaced by $D$, which may lead to incorrect detections when using threshold-based peak selection. The top plot shows the PRS-only case, where the periodic structure of the ghosts is clearly visible. 
%Instead of removing these components, our proposal groups them into clusters corresponding to each target and exploits their structure. 
Rather than attempting to remove these components at the waveform level, the proposed approach explicitly exploits their structured nature by grouping them into clusters associated with each physical target.
Based on this clustering, our method distinguishes between newly detected targets and previously tracked ones. Ambiguity-aware multilateration is then applied only to newly detected targets, while standard, ghost-free multilateration is performed for tracked targets after the matching process.% and accordingly performs multilateration with ghosts for new targets and standard ghost-free multilateration after matching for tracked targets.

 % --- First Figure ---

\begin{figure}[t]%{0.329\textwidth}
    \centering
    \includegraphics[width=\linewidth]{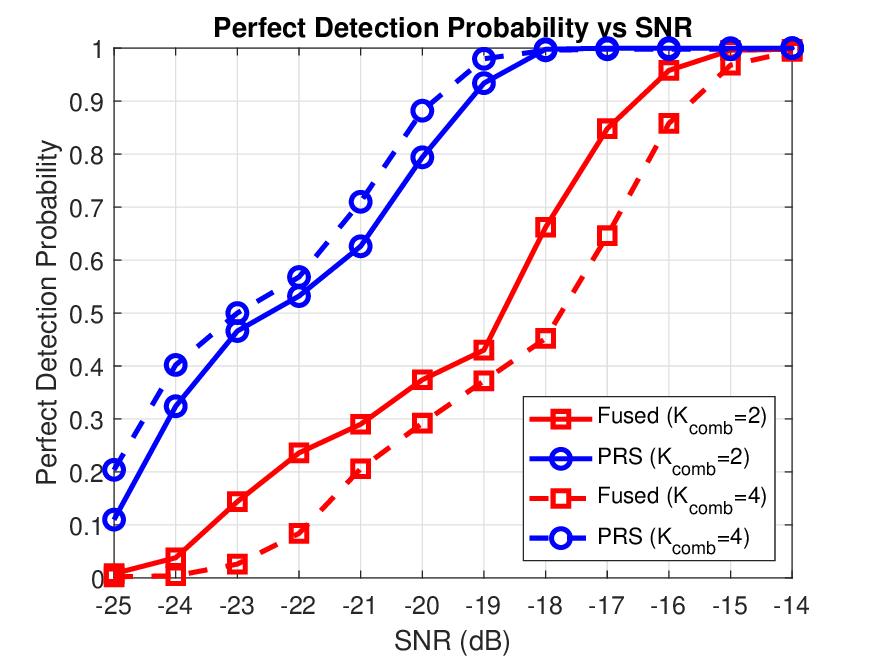}
    \caption{Comparative analysis of the perfect detection probability of our proposal and the SotA benchmark for different SNR values.}
    \label{fig:PdvsSNR}
\end{figure}

To validate the performance of the proposed method, Fig. \ref{fig:PdvsSNR} shows the perfect detection probability as a function of SNR for scenarios with a maximum of three targets. 
A perfect detection event corresponds to correctly identifying the number of targets without false positives or false negatives, and, in the PRS-only case, it requires correctly selecting the appropriate ellipse among the ambiguous candidates. Four curves are presented, corresponding to the two methods for two different comb values, $K_{\text{comb}}=2$ and $K_{\text{comb}}=4$. By averaging over 600 iterations, the results show that the PRS-only approach consistently outperforms the fused benchmark across the considered SNR values and for both $K_{\text{comb}}$ configurations. %This behavior becomes more pronounced as the number of targets increases, as observed in additional simulations with up to five targets, which are not shown here due to space limitations.

The degradation observed in the fused method is mainly due to residual peaks that persist even after combining PRS and DMRS, which can be falsely interpreted as additional targets. In contrast, the proposed method leverages the structured appearance of ghosts at distances separated by $D$, allowing it to distinguish between true targets and spurious peaks. This result highlights a key limitation of the benchmark method, which does not exploit the available side information regarding the positions of the ghost components.

\begin{figure}[t]%{0.329\textwidth}
    \centering
    \includegraphics[width=\linewidth]{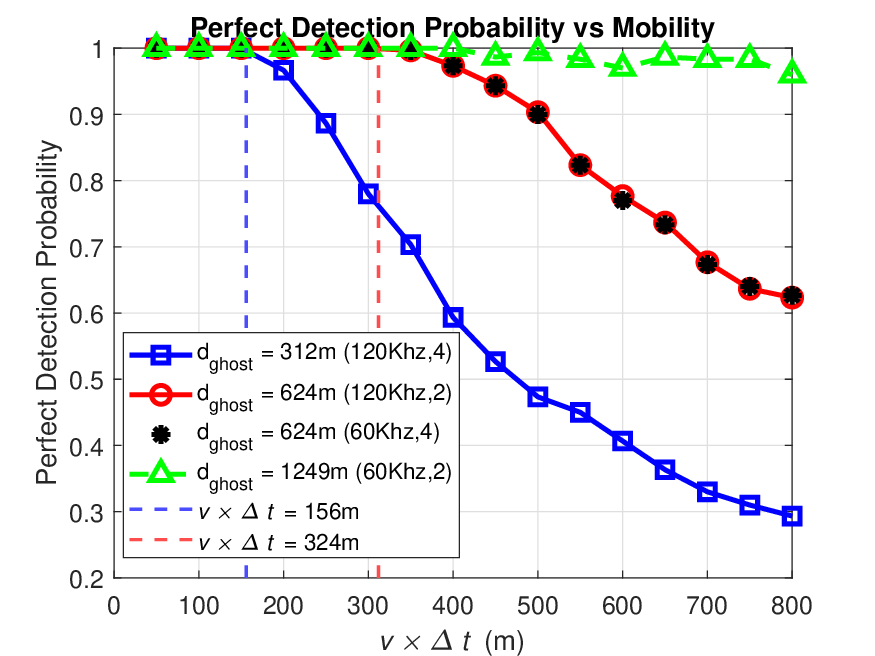}
    \caption{Analysis of the perfect detection probability vs. target displacement showing the impact of ghost spacing $D$ and the sufficient condition of Eq. (\ref{SufficientCondition}).} 
    \label{fig:PdvsV}
\end{figure}

Figure \ref{fig:PdvsV} investigates the impact of target mobility on the matching process by evaluating the error probability as a function of the displacement between consecutive observations, i.e.,  $v\Delta t$, as predicted by Eq. (\ref{SufficientCondition}).
Using the same system parameters as before and focusing on a single-target scenario in a high-SNR regime, the performance is dominated by the ability of the matching process to correctly associate current measurements with historical estimates.
 %Using the same system parameters, but focusing on a single target in a high SNR regime, we consider the case where the performance is dominated by the ability to correctly associate the current measurements with the historical estimates. 
%Considering different pairs of SCS and $K_{\text{comb}}$, four curves are presented. 
Results are shown for four configurations corresponding to different combinations of subcarrier spacing (SCS) and PRS comb size $K_{\text{comb}}$. The results are obtained by averaging over 600 iterations, where each even iteration includes a random displacement of the target position with a maximum amplitude of $v\Delta t$. The results exhibit a clear threshold behavior, in agreement with Eq. (\ref{SufficientCondition}).

Specifically, we observe that for the green curve (corresponding to $D=1249$ m), errors remain negligible over the entire range of $v\Delta t$. When reducing $D$ to 624 m (red and black curves), the probability of error remains negligible for small displacements, and then increases once the displacement exceeds the threshold around $v\Delta t = 324$ m. Further, reducing  the ghost spacing to $D=312$~m leads to a similar behavior, with the onset of errors occurring at even smaller displacements, around $v\Delta t = 156$ m. This behavior confirms the robustness of the proposed method to mobility as long as the condition in Theorem \ref{th1} is satisfied. It also highlights the trade-off in system design, where larger ghost spacing improves robustness to mobility, while smaller ghost spacing may be sufficient in low-mobility sensing scenarios.

%For small displacements, the error probability remains negligible, indicating that the proposed matching procedure is able to correctly track the targets. As the displacement increases, the error probability rises sharply, reflecting the loss of consistency between the current observations and the stored history.

%Overall, the results show that, for sufficiently high SNR and target mobility less than the threshold, the proposed method achieves near-perfect performance with no detection or matching errors. In this regime, it not only avoids the need for additional reference signals, but also outperforms the benchmark approach in terms of detection accuracy.

\begin{comment}
    
\begin{figure}[t!]

\centering % 0.524   
\includegraphics[width=0.35\textwidth]{Figures/PRS_MultilaterationFigures.tex/3Merged.eps}
\caption{Optimal sensing refresh rate $\Delta t$ versus target velocity for $K_{\text{comb}}=12$ and different $\Delta f$, satisfying the sufficient condition Eq. (\ref{SufficientCondition}) $(\Delta t \in \{20,40,80,160\}\text{ ms})$.}
%\vspace{-5mm}
\label{fig:PdvsV}
\end{figure}
\end{comment}

 % --- Second Figure ---

%\section{Conclusions and Future Works}
%\vspace{-10 pt}
\section{Conclusions}
%This paper presented a novel approach to addressing the challenges of ambiguity and false positives in 5G OFDM reference signals sensing within ISAC systems.
This paper proposed a novel PRS-only sensing framework to address the challenges of ambiguity and false detection in 5G OFDM reference signals-based ISAC systems. By using the successive estimation technique and multilateration at the fusion anchor at each new target, our proposed method reduces reliance on multiple reference signals to resolve the ambiguity problem. A sufficient condition was established to ensure correct matching, allowing the selection of appropriate sensing parameters for a given mobility scenario. Numerical results demonstrated that the proposed method can significantly reduce sensing overhead and improve detection performance while maintaining full compatibility with 3GPP-compliant 5G NR signaling. Future work will focus on extending the proposed framework to more complex and dynamic scenarios to enhance its applicability.
 
  %of the method in large-scale deployments.

 % --- Third Figure ---

%\vspace{-10 pt}
\bibliography{conf1}

@inproceedings{khosroshahi2024leveraging,
  title={Leveraging {PRS} and {PDSCH} for integrated sensing and communication systems},
  author={Khosroshahi, Keivan and Sehier, Philippe and Mekki, Sami},
  booktitle={GLOBECOM 2024},
  pages={4702--4707},
  year={2024},
  organization={IEEE}
}

@article{widdison2024review,
  title={A review of linear multilateration techniques and applications},
  author={Widdison, Eric and Long, David G},
  journal={IEEE Access},
  volume={12},
  pages={26251},
  year={2024},
  publisher={IEEE}
}

@inproceedings{al2024dynamic,
  title={Dynamic Power Allocation in {OFDM ISAC} for Time of Arrival Estimation},
  author={Al Khansa, Ali and Madhusudan, Giyyarpuram and Larue, Guillaume and Dufrene, Louis-Adrien},
  booktitle={2024 IEEE CSCN},
  pages={249--254},
  year={2024},
  organization={IEEE}
}

@inproceedings{lyu2025reference,
  title={Reference signal-based waveform design for integrated sensing and communications system},
  author={Lyu et al., Ming},
  booktitle={ICC 2025-IEEE International Conference on Communications},
  pages={6621--6626},
  year={2025},
  organization={IEEE}
}

@inproceedings{al2025adaptive,
  title={Adaptive dual-windowing strategies for multi-target detection in OFDM ISAC},
  author={Al Khansa, Ali and Bahannis, Youssef},
  booktitle={8th International Conference CommNet},
  pages={1--7},
  year={2025},
  organization={IEEE}
}

@ARTICLE{10944644,

  author={Du et al., Zhen},

  journal={IEEE Wireless Communications}, 

  title={Toward {ISAC}-Empowered Vehicular Networks: Framework, Advances, and Opportunities}, 

  year={2025},

  volume={32},

  number={2},

  pages={222-229},

  doi={10.1109/MWC.002.2300496}}

@inproceedings{al2025learning,
  title={Learning to count targets from dual-window: A CNN approach for OFDM ISAC},
  author={Al Khansa, Ali},
  booktitle={2025 8th International Conference CommNet},
  pages={1--6},
  year={2025},
  organization={IEEE}
}

@misc{3gpp2020nr,
  title={{NR}; Physical layer procedures for data},
  author={3GPP},
  year={3rd {G}eneration {P}artnership {P}roject (3{GPP}), {T}echnical {S}pecification ({TS}) 38.214, 2024, v18.2.0},
  publisher={3rd {G}eneration {P}artnership {P}roject (3{GPP}), Technical Specification ({TS})}}

@article{hakobyan2019ofdm,
  title={{OFDM-MIMO} radar with optimized nonequidistant subcarrier interleaving},
  author={Hakobyan, Gor and Ulrich, Michael and Yang, Bin},
  journal={IEEE Transactions on Aerospace and Electronic Systems},
  volume={56},
  number={1},
  pages={572--584},
  year={2019},
  publisher={IEEE}
}

@inproceedings{golzadeh2024joint,
  title={Joint Sensing and {UE} Positioning in {5G-6G: PRS} Range Estimation with Suppressed Ambiguity},
  author={Golzadeh et al., Moeinreza},
  booktitle={2024 IEEE Radar Conference (RadarConf24)},
  pages={1--6},
  year={2024},
  organization={IEEE}
}

@inproceedings{khosroshahi2024doppler,
  title={Doppler Ambiguity Elimination Using {5G} Signals in Integrated Sensing and Communication},
  author={Khosroshahi, Keivan and Sehier, Philippe and Mekki, Sami},
  booktitle={2024 IEEE 100th VTC2024-Fall)},
  pages={1--6},
  year={2024},
  organization={IEEE}
}

@article{wei2024multiple,
  title={Multiple reference signals collaborative sensing for integrated sensing and communication system towards {5G-A} and {6G}},
  author={Wei et al., Zhiqing},
  journal={IEEE Transactions on Vehicular Technology},
  year={2024},
  publisher={IEEE}
}

@inproceedings{strinati2024distributed,
  title={Distributed intelligent integrated sensing and communications: The {6G-DISAC} approach},
  author={Strinati et al., Emilio Calvanese},
  booktitle={2024 Joint European Conference on Networks and Communications \& 6G Summit (EuCNC/6G Summit)},
  pages={392--397},
  year={2024},
  organization={IEEE}
}

@phdthesis{braun2014ofdm,
  title={{OFDM} radar algorithms in mobile communication networks},
  author={Braun, Klaus Martin},
  year={2014},
  school={KIT}
}

@article{wei20225g,
  title={{5G PRS}-based sensing: A sensing reference signal approach for joint sensing and communication system},
  author={Wei et al., Zhiqing},
  journal={IEEE Transactions on Vehicular Technology},
  volume={72},
  number={3},
  pages={3250--3263},
  year={2022},
  publisher={IEEE}
}

\end{document}